\documentclass{article}
\usepackage{cite}
\usepackage{soul}
\usepackage{amsmath,amssymb,amsfonts}
\usepackage{algorithmic}
\usepackage{graphicx}
\usepackage{algorithm,algorithmic}
\usepackage{hyperref}
\usepackage{multirow}
\usepackage{textcomp}
\def\BibTeX{{\rm B\kern-.05em{\sc i\kern-.025em b}\kern-.08em
    T\kern-.1667em\lower.7ex\hbox{E}\kern-.125emX}}
\begin{document}
\title{Volumetric medical image segmentation through dual self-distillation in U-shaped networks}
\author{Soumyanil Banerjee, Nicholas Summerfield, Ming Dong, \\ Carri Glide-Hurst}

\maketitle

\begin{abstract}
U-shaped networks and its variants have demonstrated exceptional results for medical image segmentation. In this paper, we propose a novel dual self-distillation (DSD) framework in U-shaped networks for volumetric medical image segmentation. DSD distills knowledge from the ground-truth segmentation labels to the decoder layers. Additionally, DSD also distills knowledge from the deepest decoder and encoder layer to the shallower decoder and encoder layers respectively of a single U-shaped network. DSD is a general training strategy that could be attached to the backbone architecture of any U-shaped network to further improve its segmentation performance. We attached DSD on several state-of-the-art U-shaped backbones, and extensive experiments on various public 3D medical image segmentation datasets (cardiac substructure, brain tumor and Hippocampus) demonstrated significant improvement over the same backbones without DSD. On average, after attaching DSD to the U-shaped backbones, we observed an increase of 2.82\%, 4.53\% and 1.3\% in Dice similarity score, a decrease of 7.15 mm, 6.48 mm and 0.76 mm in the Hausdorff distance, for cardiac substructure, brain tumor and Hippocampus segmentation, respectively. These improvements were achieved with negligible increase in the number of trainable parameters and training time. Our proposed DSD framework also led to significant qualitative improvements for cardiac substructure, brain tumor and Hippocampus segmentation over the U-shaped backbones. The source code is publicly available at \url{https://github.com/soumbane/DualSelfDistillation}.
\end{abstract}

\section{Introduction}
\label{sec:introduction}
Deep learning algorithms have proved to be extremely powerful for Artificial Intelligence (AI) tasks with applications in Computer Vision, Medical Image Processing, Natural Language Processing, Robotics and many other areas \cite{deeplearningnature, deeplearningappls}. These algorithms learn the features of the input data in an end-to-end fashion, thereby making them extremely useful for automated pattern recognition tasks. Convolutional Neural Networks (CNNs) have been a prominent deep learning tool in numerous computer vision applications such as image classification \cite{he2016deep} and real time object detection \cite{ren2015faster}. CNNs have been widely used for many biomedical imaging tasks \cite{banerjee2020deep,jeong2021deep,mortazi2017multi} and have also given superior results with various medical imaging modalities over traditional machine learning methods \cite{nie2016estimating,banerjee2020deep,moeskops2016deep}.  

Semantic image segmentation is a challenging computer vision task where the goal is to classify each and every pixel of the image and separate a group of pixels from the rest of the image with the help of a segmentation mask that belongs to a particular class. CNNs have given promising results for semantic segmentation with fully convolutional networks \cite{long2015fully, ultrasound_seg, multistage_FCN} which perform the pixel-wise classifcation in an end-to-end manner. 

Accurate delineation of clinically relevant structures from radiology images is a crucial step in medical image analysis. However, manual delineation by a trained physician is a time-consuming task that may be prone to errors, potentially adversely impacting patient outcomes. Thus, implementing robust methods for automated segmentation in medical images are needed to facilitate a more rapid clinical workflow \cite{masood2015survey, FC_CRF}. CNNs have also proven to be extremely useful in performing medical image segmentation, which is a very important and challenging task in medical image analysis \cite{lei2020medical,wang2022medical}. One of the breakthrough algorithms which produced state-of-the-art results for end-to-end 2D and 3D medical image segmentation task is the U-Net \cite{ronneberger2015u} and the 3D U-Net \cite{cciccek20163d}, respectively. These U-shaped architectures consist of a CNN-based contracting encoder to capture the context of the input image and a CNN-based expanding decoder to localize the object in the image. Skip-connections are included between the encoder and decoder to concatenate the feature maps from encoder layers to the corresponding decoder layers. These skip-connections allow U-Nets to use the fine-grained details learned from the encoder blocks and construct a localized image in the decoder blocks. A recent variation of the CNN-based U-Net is the nnU-Net \cite{isensee2021nnu}, which uses anisotropic kernel sizes and strides for each layer of encoder and decoder along with automatic selection of the preprocessing and postprocessing steps. nnU-Net has provided state-of-the-art results for several medical image segmentation tasks.

Transformers are a class of deep neural networks that introduced the multi-head self-attention mechanism which can capture the relationship between every pair of words in a sentence in a parallel manner \cite{vaswani2017attention}. Recent transformer variants such as BERT \cite{devlin2018bert} and large language models (LLMs) such as GPT-3 \cite{brown2020language}, have been used for numerous applications such as machine translation and speech recognition. The self-attention mechanism of transformers has been introduced in computer vision by vision transformers \cite{dosovitskiy2020image}, which have shown promising results for image classification. Vision transformers (ViTs) eliminate problem faced by CNNs where the small receptive field of the CNN kernel can only capture local information from the image. The self-attention mechanism help ViTs capture the long-range dependencies between different regions of the image. ViTs have been used in the encoder path of U-shaped networks along with a CNN-based decoder path by UNETR \cite{hatamizadeh2022unetr} to yield state-of-the-art results for medical image segmentation tasks. The efficiency of ViT in the encoder path of UNETR was further improved with the inclusion of a swin transformer \cite{swintransformer} in the encoder path as done by Swin UNETR \cite{hatamizadeh2022swin, swinunetrpretrain} along with a CNN-based decoder path. Swin UNETR was more efficient than UNETR due to the shifted window mechanism of swin transformers \cite{swintransformer}.

Ensemble methods such as bagging and boosting improve the performance of machine learning models by training multiple models and combining the output of these models \cite{ensembleml}. Ensemble methods have been used in deep learning by training multiple models with different random initialization and then combining the outputs of these models during the inference phase to improve the prediction performance. However, prediction with multiple models during inference phase is time consuming, making the process infeasible for real-time applications especially on edge devices such as smart phones. 

This led researchers to develop a method known as knowledge distillation, which is the process by which a large pre-trained model acts as a teacher network, transferring its knowledge to a smaller, lightweight model that acts as a student network during its training \cite{hinton2015distilling,allen2020towards,sadowski2015deep}. Knowledge distillation was first proposed to reduce the inference time by using a lightweight model while maintaining similar accuracy to the large pre-trained model. With knowledge distillation, smaller lightweight models could be also deployed on edge devices. Recently, knowledge distillation has been used to improve the performance of lightweight networks for semantic segmentation \cite{Liu_2019_CVPR} including medical image segmentation tasks \cite{qin2021efficient}. However, knowledge distillation still requires the training of a large network in order to distill the knowledge to a smaller network for different prediction tasks. 

The need of transferring knowledge from a large teacher network to a smaller student network was later eliminated by the process of self-distillation \cite{zhang2019your}. Self-distillation uses the deepest layer of a single model to act as the teacher network in order to distill knowledge to the shallower layers of the same model, acting as the student network. Since self-distillation uses a single model, it is much more efficient when compared to ensemble methods. Additionally, self-distillation is much faster to train as it is not necessary to train a large network as is required by knowledge distillation. Self-distillation has been applied for numerous computer vision tasks such as image classification \cite{zhang2019your,zhang2021self} and object detection \cite{zhang2022lgd}. A rigorous theoretical explanation of ensemble, knowledge distillation and self-distillation for image classification was recently provided in \cite{allen2020towards}. More recently, the self-distillation process was applied to a transformer based TransUNet \cite{chen2021transunet}, along with multi-scale fusion blocks to perform 3D medical image segmentation \cite{wang2023missu}. However, this method was limited to the specific TransUNet architecture.

In this paper, we address the question of whether we can train a single U-shaped network and leverage the learned representations of the deeper encoder and decoder network layers to improve the prediction performance of medical image segmentation. Here, we propose a novel volumetric dual self-distillation (DSD) framework that could independently be attached to any U-shaped image segmentation backbone. In DSD, the deepest encoder and decoder of the U-shaped backbone act as the teacher networks for the shallower encoders and decoders which act as student networks. We found that the deepest encoder which is at the bottom of the contracting encoder path of the U-shaped network contains more contextual information compared to the shallower encoder layers. Similarly, the deepest decoder which is at the top of the expanding decoder path of the U-shaped network contains more semantic information than the shallower decoder layers. Thus, in our DSD framework, this information or knowledge distills in a bottom-up manner on the encoder side and in a reverse top-down manner on the decoder side of a U-shaped backbone. Additionally, DSD also includes the distillation of knowledge from the Ground-truth Labels (GT Labels) to the decoder layers of the U-shaped network, which is a process known as deep supervision in medical image segmentation \cite{dou20173d}. The self-distillation process of DSD not only allows intermediate decoder and encoder representations to be optimized and aligned with the final representation of the U-shaped network, but also allows the hidden knowledge to flow from the deepest decoder and encoder layers to the shallower decoder and encoder layers respectively, and hence benefits from both regularization and generalization. In this manner, DSD leverages the benefits of deep supervision by overcoming optimization difficulties and achieving faster convergence \cite{dou20173d}.

Our major contributions are summarized as follows: 

\begin{enumerate}
    \item We proposed self-distillation in U-shaped networks for volumetric medical image segmentation. Our novel design of DSD between encoders and decoders could be generalized to any U-shaped segmentation backbone.
    \item The self-distillation process in our DSD framework offers a more general training approach than deep supervision, which can be considered a special case of our proposed framework. Thus, DSD leverages the benefits of deep supervision and can alleviate the problem of vanishing gradients and promote faster convergence, thereby improving the segmentation performance of any U-shaped backbone.
    \item We performed extensive experiments on various public 3D medical image segmentation datasets (one with cardiac substructures, one with brain tumors and another segmenting the Hippocampus), with DSD attached to several state-of-the-art U-shaped backbones (e.g., nnU-Net, UNETR, VNet and Swin UNETR) and demonstrated significant quantitative and qualitative improvements over those backbones with negligible increase in model parameters and training time.
\end{enumerate}

A preliminary version of this work was presented at ISBI 2024 \cite{DSD_ISBI}. In this paper, we have thoroughly expanded the theoretical explanation of DSD as a regularization term, analyzed the effect of network parameters, and introduced new experiments with additional U-shaped networks. We also compared our framework with state-of-the-art methods and performed extensive ablation studies to validate our design choices. The rest of the paper is organized as follows: Section \ref{sec:method} describes the data set and the details of the DSD framework. Section \ref{sec:results} describes the experimental setup and the results of our experiments. Section \ref{sec:discussion} presents discussion and future applications of our framework. Lastly, Section \ref{sec:conclusion} presents our conclusion.

\section{Methodology}
\label{sec:method}
\begin{figure*}[ht]
\centering
\includegraphics[width=\textwidth]{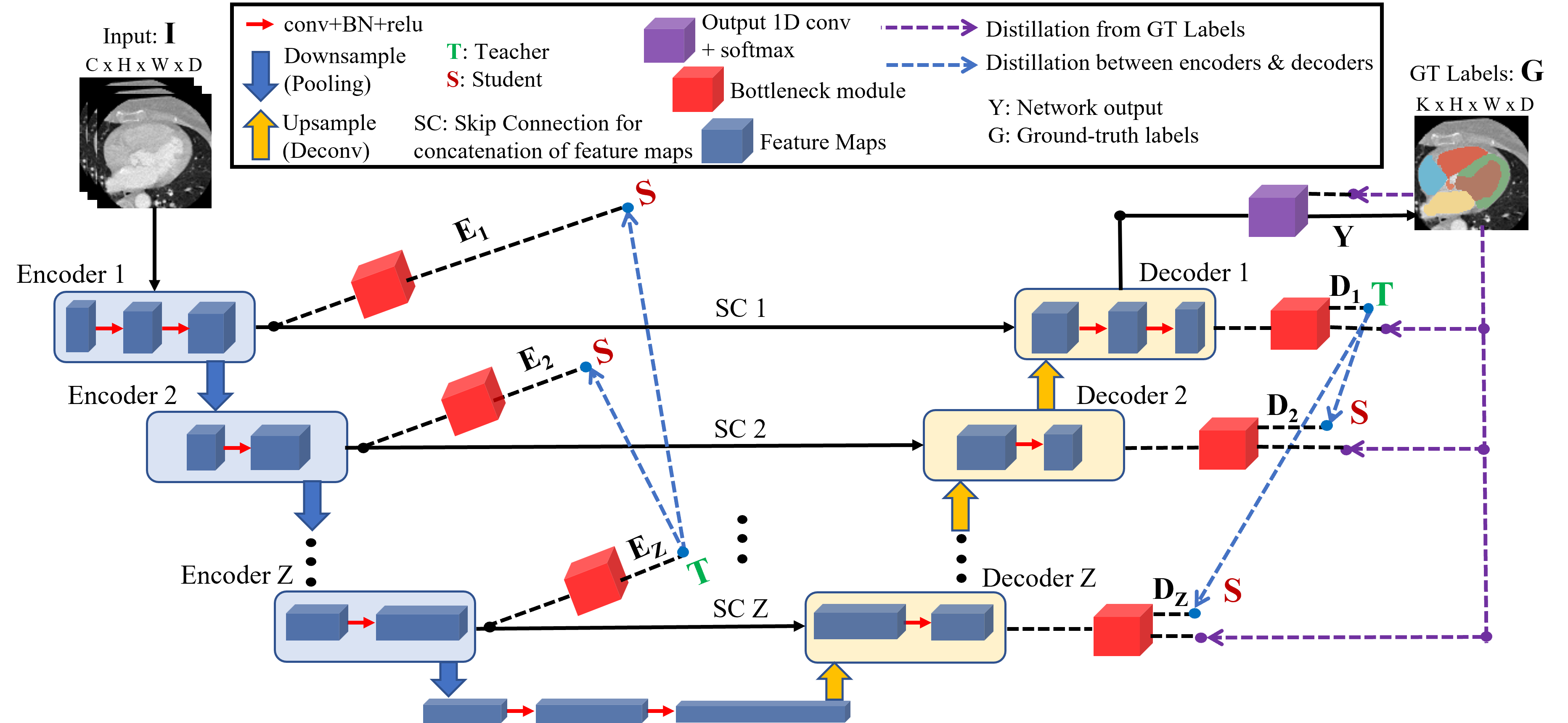}
\caption{Self-distillation demonstrated with an U-shaped network for volumetric medical image segmentation. $Y$ and $G$ indicate the softmax of network output and one-hot encoded GT Labels, respectively. $E_i \vert_{i=1}^{Z}$ and $D_i \vert_{i=1}^{Z}$ denote the output of the bottleneck module on the encoder and decoder side, respectively. $T$ and $S$ denote the teacher and student probability distributions, respectively. All dashed lines shown are only used during training and removed during inference. The input is shown as a stack of images to indicate a 3D CT volume, and the output shows one-slice of the GT Labels overlayed on a 2D CT slice for a clear visualization of the different segmentation classes.}
\label{fig:figure1}
\end{figure*}

In the following sections, we provide a detailed explanation of the datasets, the preprocessing steps used and the components of our proposed DSD framework as shown in Fig. \ref{fig:figure1}.

\subsection{Datasets}
\label{subsec:dataset}
We used three publicly available datasets to train and evaluate our proposed DSD framework. These datasets were chosen as they cover different organs of the human body (whole heart, brain tumor and hippocampus), and they were acquired with different techniques such as Computed Tomography (CT), T1-weighted Magnetic Resonance Imaging (MRI) and multi-modal (T1w, T2w, FLAIR and T1c) multi-site MRI. Evaluating with these datasets would ensure that our proposed approach could be applied to various real-world radiology images involving different anatomies of the human body.

\noindent \textbf{Cardiac dataset (MMWHS)} - This comprised of high resolution 3D CT angiography (CTA) data of 20 patients from the Multi-Modal Whole Heart Segmentation (MMWHS) dataset \cite{luo2022mathcal,zhuang2018multivariate,zhuang2016multi}. Each 3D image were of variable sizes along the three dimensions with a single input channel for the CT modality. The GT Labels consisted of 7 foreground classes of cardiac substructures and a background class. The foreground classes are: (i) Myocardium (MYO), (ii) Left Atrium (LA), (iii) Left Ventricle (LV), (iv) Right Atrium (RA), (v) Right Ventricle (RV), (vi) Ascending Aorta (AA), (vii) Pulmonary Artery (PA). We split the data into a training and validation set of 16 and 4 patients respectively. Then, we performed a 5-fold cross validation over the entire dataset.

We performed a set of preprocessing steps on the dataset prior to using them for training and evaluation with our network. The sequence of preprocessing steps used prior to training are: (i) We resampled all of the input images and GT Labels to have a voxel dimension of 1.5 mm along the coronal and sagittal direction and 2 mm along the axial direction. Bilinear interpolation was used to resample the images and nearest neighbor interpolation was used to resample the GT Labels. (ii) The orientation of all the input images and GT Labels were changed to have a uniform orientation of ``Right'', ``Anterior'' and ``Superior'' for the coronal, sagittal and axial directions respectively. (iii) We scaled the intensity range of all the input images between 0 and 1 and then normalized each dimension of the image separately using the z-score normalization. (iv) To generate augmentations and have a uniform size for all the images and GT Labels, we randomly cropped out two 3D volumes of size (96,96,96) with the foreground voxel as the center. (v) These images and labels were then randomly flipped along all the three axis of the 3D volume with a probability of 10\%. (vi) The images and labels were randomly rotated by 90 degrees with a probability of 10\% along the coronal-sagittal plane. (vii) The intensity of the input images were randomly shifted between (-0.1,0.1) with a probability of 50\%.

For evaluation purposes, only the preprocessing steps (i), (ii) and (iii) were performed as the steps (iv)-(vii) are only used to perform data augmentations needed for training our network with the MMWHS dataset. 

\noindent \textbf{Brain tumor dataset (MSD-BraTS)} - The brain tumor segmentation (BraTS) task \cite{bakas2017advancing,bakas2018identifying,BraTS_Dataset} was used from the Medical Segmentation Decathlon (MSD) dataset \cite{antonelli2022medical,simpson2019_msd}. This dataset comprised of a combination of BraTS 2016 and 2017 datasets. The segmentation task comprised of 484 patients having multi-modal multi-site MRI data. Each 3D image consisted of 4 channels for the multi-modal MRI input (FLAIR, T1w, T1gd and T2w). The GT Labels consisted of 3 foreground classes and a background class. The foreground classes are: (i) Whole tumor (WT), (ii) Enhancing tumor (ET) and (iii) Tumor core (TC). We took the approach adopted by UNETR \cite{hatamizadeh2022unetr} and split the data into a training, validation and testing set of 388, 72 and 24 patients respectively.

The sequence of preprocessing steps used prior to training are: (i) The orientation of all the input images and GT Labels were changed to have a uniform orientation of ``Right'', ``Anterior'' and ``Superior'' for the coronal, sagittal and axial directions respectively. (ii) To generate augmentations and have a uniform size for all the images and GT Labels, we randomly cropped out two 3D volumes of size (128,128,128) with the foreground voxel as the center. (iii) These images and labels were then randomly flipped along all the three axis of the 3D volume with a probability of 50\%. (iv) The images and labels were randomly rotated by 90 degrees with a probability of 10\% along the coronal-sagittal plane. (v) The intensity of the input images were shifted between (-0.1,0.1) with a probability of 100\%.

For evaluation purposes, only the preprocessing step (i) was performed as the steps (ii)-(v) are only used to perform data augmentations needed for training our network with the MSD-BraTS dataset.

\noindent \textbf{Hippocampus Dataset (Hippocampus)} - The Hippocampus segmentation task was used from the Medical Segmentation Decathlon (MSD) dataset \cite{antonelli2022medical,simpson2019_msd}. The segmentation task comprised of 260 patients with 3D T1-weighted MRI images. Each volume was annotated by using 2 labels for hippocampus and parts of the subiculum. Hence, the GT Labels consisted of 2 foreground classes and a background class. The foreground classes are: (i) Anterior, and (ii) Posterior. We divided this dataset into 200 patients for training and validation and 60 patients for held-out testing.

The sequence of preprocessing steps used prior to training are: (i) The orientation of all the input images and GT Labels were changed to have a uniform orientation of ``Right'', ``Anterior'' and ``Superior'' for the coronal, sagittal and axial directions respectively. (ii) We scaled the intensity range of all the input images between 0 and 1. (iii) To generate augmentations and have a uniform size for all the images and GT Labels, we randomly cropped out 16 3D volumes of size (32,32,32) with the foreground voxel as the center. (iv) These images and labels were then randomly flipped along all the three axis of the 3D volume with a probability of 50\%. (v) The images and labels were randomly rotated by 90 degrees with a probability of 10\% along the coronal-sagittal plane. (vi) The intensity of the input images were shifted between (-0.1,0.1) with a probability of 100\%.

For evaluation purposes, only the preprocessing steps (i) and (ii) were performed as the steps (iii)-(vi) are only used to perform data augmentations needed for training our network with the Hippocampus dataset.

It should be noted that during training, we are performing the random rotations only at 90-degree angles and not arbitrary angles. This is an important augmentation strategy since it simulates the CT or MRI acquisitions where the acquisition could be along any of the axial, coronal or sagittal planes. Hence, by randomly rotating the CT or MRI volumes by 90-degrees, we could inform the model about possible image acquisition strategies such that the model could generalize well during inference time to volumes that were acquired along the axial, coronal or sagittal planes. For evaluation purposes, the steps we performed are the standard steps followed by previous studies in the medical image segmentation literature (\cite{ronneberger2015u}, \cite{hatamizadeh2022unetr}, \cite{milletari2016v}, \cite{hatamizadeh2022swin}). These evaluation steps bring all volumes to the same orientation (“RAS” for our case) which is a crucial step necessary for all algorithms. Moreover, the uniform voxel dimension and scaled intensity are the least computationally intensive steps that could be easily performed on a new test image. Hence, all previous algorithms used them, and we followed the same approach in this paper for all baseline algorithms with and without DSD.

This research study was conducted retrospectively using human subject data made available in open access for MMWHS \cite{luo2022mathcal,zhuang2018multivariate,zhuang2016multi}, MSD-BraTS \cite{bakas2017advancing,bakas2018identifying,BraTS_Dataset} and Hippocampus \cite{antonelli2022medical,simpson2019_msd} datasets. Ethical approval was not required as confirmed by the license attached with the open access data.

\subsection{U-shaped backbone}
\label{subsec:ushapednet}
The U-shaped backbone maps an input image $I$ ($I \in \mathbb{R}^{C \times H \times W \times D}$ with $H$, $W$, $D$ denoting the height, width and depth of a 3D input image, respectively and $C$ denoting number of imaging modalities/sequences such as CT or multi-modal MRI) to the GT Labels $G$ ($G \in \mathbb{R}^{K \times H \times W \times D}$ with $K$ classes).

The most common loss function used by a U-shaped network \cite{hatamizadeh2022swin,hatamizadeh2022unetr} is the Dice Cross-Entropy (CE) loss ($L_{DCE}$), which is a compound loss function defined for 3D multi-class image segmentation as follows:
\begin{equation}
\small
\begin{aligned}
L_{DCE}^{Y} &= L_{Dice}^{Y} + L_{CE}^{Y} \\
            &= 1 - \frac{2}{K} \sum_{k=1}^{K} \frac{\sum_{p=1}^N G_{p,k}Y_{p,k}}{\sum_{p=1}^N G_{p,k}^2 + \sum_{p=1}^N Y_{p,k}^2} \\
            & - \frac{1}{N} \sum_{p=1}^N \sum_{k=1}^{K} G_{p,k} \log Y_{p,k}    
\end{aligned}
\label{equation:DiceCE1}
\end{equation}
where, $G_{p,k} \in \mathbb{R}^{K \times H \times W \times D}$ denotes the one-hot encoded GT Labels and $Y_{p,k} \in \mathbb{R}^{K \times H \times W \times D}$ denotes the probability (softmax) output of the network for class $k$ at pixel $p$ (here $p$ denotes the 3D pixel position, i.e., one 3D coordinate of ($H,W,D$)). $N = H*W*D$ denotes the total number of pixels in input image $I$ and GT Labels $G$. The Dice loss $L_{Dice}^Y$ measures the pixel-wise similarity while the cross-entropy loss $L_{CE}^Y$ measures the pixel-wise difference in distributions between the network output $Y$ and GT Labels $G$. This loss is back-propagated to the network to update the weights of the encoders and decoders.

\subsection{Bottleneck Module}
\label{subsec:bottlenecks}
The bottleneck module shown by the red colored boxes in Fig. \ref{fig:figure1} constitutes an integral component of our proposed framework, which converts the feature maps $F$ ($F \in \mathbb{R}^{K' \times H' \times W' \times D'}$) obtained from different encoder and decoder layers to a probability distribution $P$ ($P \in \mathbb{R}^{K \times H \times W \times D}$) of the same shape as the network output $Y$. For the feature maps $F$, $K', H', W'$ and $D'$ denote the number of channels, height, width and depth respectively at a given layer, and they vary depending on the position of the encoder (Encoder $i \vert_{i=1}^{Z}$) and decoder (Decoder $i \vert_{i=1}^{Z}$) layers.

The bottleneck module consist of three layers: (i) 1D convolution layer: this layer changes the number of channels of feature maps obtained from the encoder and decoder layers to match the number of output classes $K$ (from $F \in \mathbb{R}^{K' \times H' \times W' \times D'}$ to $F' \in \mathbb{R}^{K \times H' \times W' \times D'}$), (ii) Deconvolution layer: this layer upsamples the feature maps obtained from the 1D convolution layer to generate logits ($L$) that match the dimension of the output $Y$ (from $F' \in \mathbb{R}^{K \times H' \times W' \times D'}$ to $L \in \mathbb{R}^{K \times H \times W \times D}$), (iii) Softmax layer: this layer converts the logits $L$ ($L \in \mathbb{R}^{K \times H \times W \times D}$) to soft labels which is a probability distribution $P$ of the same shape as $L$, given by: 

\begin{equation}
\small
\begin{aligned}
P_{p,k} = \frac{\exp^{L_{p,k}/\tau}}{\sum_{j=1}^K \exp^{L_{p,j}/\tau}}
\end{aligned}
\label{equation:softlabelsgenerator}
\end{equation}
where, $P_{p,k}$ denote the soft labels generated for class $k$ at pixel $p$, $L$ denotes the logits, $\tau$ ($\tau > 1$) denotes the temperature to generate the soft labels, and $p \in N$ and $k \in K$ are indices for pixels and classes, respectively.

\subsection{U-shaped backbone with Dual Self-distillation (DSD)}
\label{subsec:ushapednetselfdistil}
We propose a novel DSD framework for U-shaped backbones as shown by the purple and blue dashed arrows in Fig. \ref{fig:figure1}. Our DSD framework consists of two main components. 

\noindent (i) \textbf{Distillation from GT Labels}: the first part (purple dashed arrows in Fig. \ref{fig:figure1}) is the distillation of knowledge from the GT Labels $G$ to each decoder of the U-shaped network. This process is known as deep supervision in medical image segmentation \cite{dou20173d}. For our DSD framework, we calculate the Dice Cross-Entropy (DCE) loss (Eq. \ref{equation:DiceCE1}) between each decoder layer's softmax output $D_i \vert_{i=1}^{Z}$ and the GT Labels $G$. This loss is defined as:
\begin{equation}
\small
\begin{aligned}
L_{DS} &= L_{DCE}^{Y} + \eta \sum_{i=1}^{Z} L_{DCE}^{D_i} \\
       &= L_{DCE}^{Y} + \eta \sum_{i=1}^{Z} \left(1 - \frac{2}{K} \sum_{k=1}^{K} \frac{\sum_{p=1}^N G_{p,k}D_{i_{p,k}}}{\sum_{p=1}^N G_{p,k}^2 + \sum_{p=1}^N D_{i_{p,k}}^2} \right) \\
       &- \eta \sum_{i=1}^{Z} \left(\frac{1}{N} \sum_{p=1}^N \sum_{k=1}^{K} G_{p,k} \log D_{i_{p,k}} \right)
\end{aligned}
\label{equation:deepsuperDiceCE1}
\end{equation}
where, $L_{DS}$ denotes the deep supervision loss, $Z$ denotes the number of decoders in the U-shaped architecture, $L_{DCE}^{D_i}$ denotes the Dice cross-entropy loss between $i^{th}$ decoder and $G$, and $\eta$ denotes the coefficient that controls the amount of supervision from $G$ to $D_i \vert_{i=1}^{Z}$.

\noindent (ii) \textbf{Distillation between encoder and decoder layers:} the second part (blue dashed arrows in Fig. \ref{fig:figure1}) is the distillation of knowledge between encoder and decoder layers of the U-shaped network. On the encoder side, the deepest encoder (Encoder Z) forms the teacher network to the shallower encoders (Encoder 1, 2, ..., (Z-1)) which form the student networks. We reverse the order of teacher and student on the decoder side due to the deconvolution operation. Hence, the deepest decoder (Decoder 1) forms the teacher network to the shallower decoders (Decoder 2, 3, ..., Z) which form the student networks. For all the teacher-student pairs in the encoders and decoders of the U-shaped network, we compute the pixel-wise Kullback–Leibler (KL) divergence \cite{Joyce2011} between the output probability distributions (softmax) of teacher and student as follows:
\begin{equation}
\small
\begin{aligned}
L_{KL} &= \alpha_1 \sum_{i=1}^{Z-1} D_{KL}(E_i,E_Z) + \alpha_2 \sum_{i=2}^{Z} D_{KL}(D_i,D_1) \\
&= \alpha_1 \sum_{i=1}^{Z-1} \left(\frac{1}{N} \sum_{p=1}^{N} \sum_{k=1}^{K} E_{Z_{p,k}} \log \frac{E_{Z_{p,k}}}{E_{i_{p,k}}} \right) \\
&+ \alpha_2 \sum_{i=2}^{Z} \left(\frac{1}{N} \sum_{p=1}^{N} \sum_{k=1}^{K} D_{1_{p,k}} \log \frac{D_{1_{p,k}}}{D_{i_{p,k}}} \right)
\end{aligned}
\label{equation:KL_loss1}
\end{equation}
where $D_{KL}(P^S,P^T)$ is the KL divergence between student ($P^S$) and teacher ($P^T$) probability distributions, $E_i$ and $D_i$ are the $i^{th}$ shallow encoder and decoder's (student's) softmax output ($P^S$) respectively, $E_Z$ and $D_1$ are the deepest encoder and decoder's (teacher's) softmax output ($P^T$), respectively, $Z$ is the number of encoders and decoders, $K$ denotes the number of classes of the GT Labels and $N$ denotes the total number of pixels.

We define our proposed DSD loss $L_{DSD}$ as follows:
\begin{equation}
\small
\begin{aligned}
L_{DSD} &= L_{DCE}^{Y} + \eta \sum_{i=1}^{Z} L_{DCE}^{D_i} \\
&+ \alpha_1 \sum_{i=1}^{Z-1} D_{KL}(E_i,E_Z) + \alpha_2 \sum_{i=2}^{Z} D_{KL}(D_i,D_1)
\end{aligned}
\label{equation:KL_loss_final}
\end{equation}
where $\alpha_1$ and $\alpha_2$ denote the coefficients that controls the amount of self-distillation between the encoder and decoder layers, respectively. Note that our DSD framework is reduced to deep supervision when $\alpha_1,\alpha_2 = 0$.

Therefore, our objective function is to minimize the loss function in Eq. \ref{equation:KL_loss_final} as follows:
\begin{equation}
\small
\arg \min_{\theta} L_{DSD}
\label{equation:objectivefunc}
\end{equation}
where, $L_{DSD}$ is the DSD loss and $\theta$ are the set of parameters that minimizes the DSD loss.

Note that the training with our DSD framework (shown by dashed arrows in Fig. \ref{fig:figure1}) uses very few extra parameters (brought by the bottleneck modules) and hence performs end-to-end training without increasing the training time when compared to the backbone architectures. During inference, the DSD framework is removed and hence it takes the same inference time as the U-shaped backbones.

\section{Experiments and Results}
\label{sec:results}

\subsection{Experimental setup and implementation details}
\label{subsec:expsetup}
We attached our DSD framework on several state-of-the-art U-shaped backbones and applied it on various benchmark datasets for 3D medical image segmentation tasks, specifically whole heart, brain tumor and Hippocampus segmentation.

The following U-shaped backbones were used for our experiments:
\begin{enumerate}
    \item \textbf{UNETR \cite{hatamizadeh2022unetr}}: A vision transformer (ViT) \cite{dosovitskiy2020image} with 12 layers was used on the embedded 3D input patches, where each patch of dimension (16,16,16) was extracted from the 3D input image. The embedding dimension of the patches was 768. The feature size used to map the input channels to the intermediate output channels (after the convolution layers on the input) was 32 for MMWHS dataset, 16 and 64 for light UNETR and standard UNETR respectively for MSD-BraTS dataset and 16 for the Hippocampus dataset. The number of heads for the multi-head self-attention in ViT was 12 and the multi-layer perceptron (MLP) dimension in ViT was 3,072. A sequence representation was extracted after layers 4, 7, 10 and 12. These representations are then passed through some encoder layers and then concatenated via skip connections with the decoder layer outputs as done in U-Net \cite{ronneberger2015u} and 3D U-Net \cite{cciccek20163d}.

    \item \textbf{nnU-Net \cite{isensee2021nnu}}: The network consists of an input block to process the 3D input images, 3 downsample blocks, a bottleneck block at the bottom of the U-shaped structure and 4 upsample blocks to predict the GT Labels. For the input, downsample, bottleneck and upsample blocks, a convolution kernel of size (3,3,3) was used. For the input block, a stride of (1,1,1) was used whereas for the downsample and upsample blocks, a stride of (2,2,2) was used. For the MMWHS dataset, the number of CNN filters for the input block was 32 and the number of CNN filters for the downsample and upsample blocks were 64, 128 and 256. For the bottleneck layer, the number of filters were 320. For the MSD-BraTS and Hippocampus dataset, the number of CNN filters for the input block was 16 and the number of CNN filters for the downsample and upsample blocks were 32, 64 and 128. For the bottleneck layer, 256 filters were used.

    \item \textbf{Swin UNETR \cite{hatamizadeh2022swin}}: The encoder path consisted of 4 stages where each stage was comprised of 2 swin transformer \cite{swintransformer} blocks along with patch merging. Patches of dimension (2,2,2) were extracted from the 3D input image. The window size for the swin transformer block was (7,7,7). The feature size used to map the input channels to the intermediate output channels (after the patch partition layer) was 12 for the MSD-BraTS dataset. The number of heads for the multi-head self-attention was 3, 6, 12, 24 for stage 1, 2, 3 and 4, respectively. The encoded feature representations from the swin transformer blocks were fed to a CNN-based decoder via skip connection at multiple resolutions.

    \item \textbf{VNet \cite{milletari2016v}}: The network consists of an input block, four downsample blocks, a bottleneck block, and four upsample blocks. The input block uses a convolution with a kernel size of 5 and outputs 16 feature maps. The first downsample block outputs 32 channels, doubling the feature map size in each successive block. The bottleneck block operates on 256 feature maps before transitioning to the upsampling blocks. The upsample blocks reverse the downsampling process by successively halving the number of channels (256, 128, 64, 32) and doubling the spatial dimensions. Each upsample block concatenates the output from the corresponding downsample block to form skip connections. The final block consists of a convolutional layer with a kernel size of 5, followed by an additional convolution to reduce the channels to the number of output labels.    
\end{enumerate}

\begin{table*}[htbp]
\centering
\caption{Quantitative comparison on high-resolution cardiac CTA MMWHS dataset with UNETR and nnU-Net with and without the DSD framework. The mean and standard deviation of Dice score (\%) and HD95 (mm) over the 5-folds are shown for each cardiac sub-structure. The highest Dice score and lowest HD95 are marked in \textbf{bold}.}
\label{tab:Table1}
\resizebox{\textwidth}{!}{%
\begin{tabular}{c|cc|cc|cc|cc}
\hline
Network &
  \multicolumn{2}{c|}{UNETR \cite{hatamizadeh2022unetr}} &
  \multicolumn{2}{c|}{UNETR with DSD (Ours)} &
  \multicolumn{2}{c|}{nnU-Net \cite{isensee2021nnu}} &
  \multicolumn{2}{c}{nnU-Net with DSD (Ours)} \\ \hline
Sub-structure & Dice $\uparrow$  & HD95 $\downarrow$ & Dice $\uparrow$  & HD95 $\downarrow$ & Dice $\uparrow$  & HD95 $\downarrow$ & Dice $\uparrow$  & HD95 $\downarrow$ \\ \hline
MYO           & 81.84 $\pm$ 4.21 & 15.98 $\pm$ 12.55 & 82.25 $\pm$ 4.07 & 17.34 $\pm$ 13.81 & 84.23 $\pm$ 4.96 & 22.08 $\pm$ 16.22 & 83.50 $\pm$ 4.58 & 13.65 $\pm$ 7.27  \\ 
LA            & 86.74 $\pm$ 2.99 & 10.63 $\pm$ 7.250 & 87.65 $\pm$ 2.20 & 9.610 $\pm$ 7.120 & 88.11 $\pm$ 4.66 & 15.20 $\pm$ 2.770 & 89.96 $\pm$ 3.67 & 8.540 $\pm$ 6.47  \\ 
LV            & 83.19 $\pm$ 7.16 & 19.74 $\pm$ 9.840 & 85.24 $\pm$ 5.77 & 14.79 $\pm$ 12.60 & 88.44 $\pm$ 4.97 & 14.36 $\pm$ 9.500 & 89.51 $\pm$ 4.61 & 12.71 $\pm$ 9.65  \\ 
RA            & 70.53 $\pm$ 5.28 & 30.52 $\pm$ 16.80 & 76.10 $\pm$ 4.51 & 29.39 $\pm$ 9.200 & 79.86 $\pm$ 5.52 & 34.04 $\pm$ 13.95 & 84.94 $\pm$ 1.84 & 20.60 $\pm$ 8.97  \\ 
RV            & 76.28 $\pm$ 6.52 & 32.77 $\pm$ 16.75 & 79.43 $\pm$ 6.38 & 26.66 $\pm$ 13.56 & 81.33 $\pm$ 5.89 & 42.98 $\pm$ 13.83 & 82.52 $\pm$ 6.25 & 27.05 $\pm$ 8.93  \\ 
AA            & 79.31 $\pm$ 7.76 & 23.32 $\pm$ 8.570 & 81.95 $\pm$ 7.70 & 18.74 $\pm$ 8.390 & 86.81 $\pm$ 4.61 & 20.64 $\pm$ 5.900 & 91.18 $\pm$ 1.93 & 8.900 $\pm$ 6.28  \\ 
PA            & 69.74 $\pm$ 8.10 & 31.37 $\pm$ 8.100 & 73.90 $\pm$ 5.15 & 25.26 $\pm$ 7.160 & 76.04 $\pm$ 8.48 & 34.91 $\pm$ 10.08 & 83.80 $\pm$ 3.73 & 15.12 $\pm$ 9.14  \\ \hline
mean $\pm$ std &
  78.24 $\pm$ 3.85 &
  23.48 $\pm$ 9.000 &
  80.94 $\pm$ 3.62 &
  20.26 $\pm$ 8.230 &
  83.55 $\pm$ 4.53 &
  26.31 $\pm$ 6.980 &
  \textbf{86.49} $\pm$ \textbf{2.28} &
  \textbf{15.23} $\pm$ \textbf{5.11} \\ \hline
\end{tabular}
}
\end{table*}

\begin{figure*}[ht]
\centering
\includegraphics[width=0.99\textwidth]{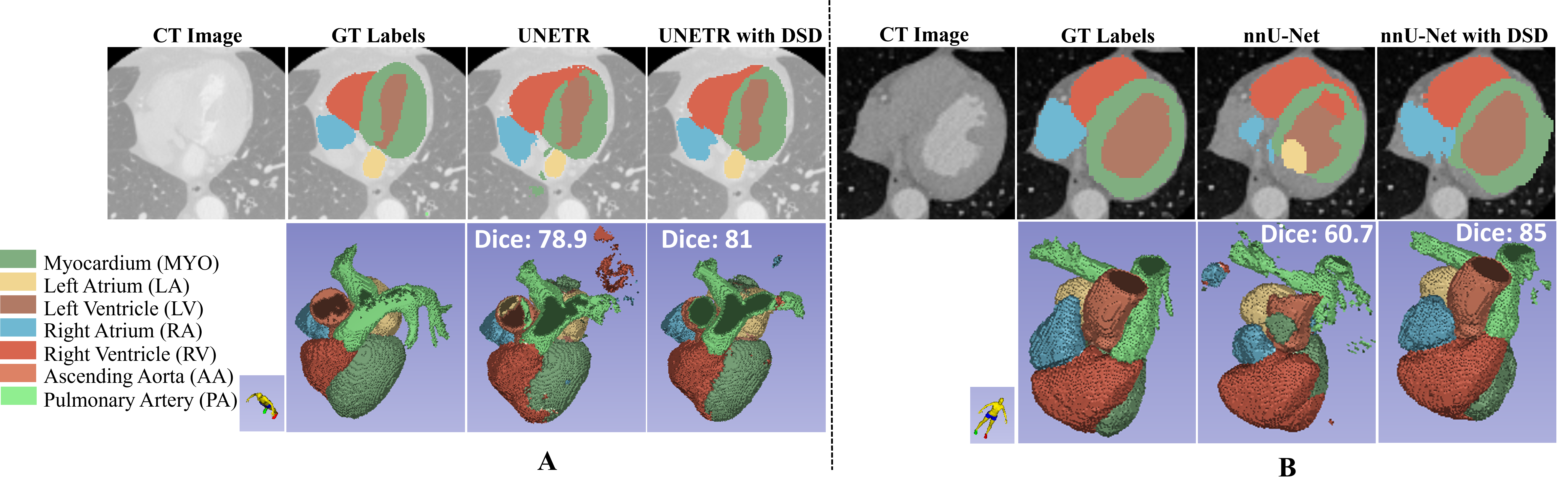}
\caption{Qualitative comparison of an axial slice and 3D volumes, with GT Labels (on CTA) and predictions with (A) UNETR and (B) nnU-Net, highlighting the improved segmentations (shown by Dice score (\%)) with our proposed DSD framework. The Dice score (\%) is for the full 3D volume of a patient belonging to the validation set.}
\label{fig:figure2}
\end{figure*}

We attached our DSD framework on the above U-shaped backbones as shown in Fig. \ref{fig:figure1} with the same configurations as described above. These U-shaped backbones were selected because they have recently shown promising and state-of-the-art results for medical image segmentation. For all experiments, the training was performed including the background class but evaluated only on the foreground classes. For the MMWHS dataset, we trained for 600 epochs with UNETR and 300 epochs with nnU-Net, both with an initial learning rate of 0.001 with step learning rate decay. For the MSD-BraTS dataset, we trained for 300 epochs with nnU-Net with an initial learning rate of 0.001 with step learning rate decay and 325 epochs with UNETR and Swin UNETR with a learning rate of 0.0001 without any learning rate decay. For the Hippocampus dataset, we trained for 300 epochs for VNet, UNETR and nnU-Net with a learning rate of 0.0001 without any learning rate decay. All DSD experiments were performed with $\eta = 1$ and $\alpha_1,\alpha_2 = 1$. The temperature ($\tau$) used to generate the soft labels in our DSD experiments was 3. These hyperparameters were empirically decided based on the performance on the validation set. The batch size used for all our experiments was 1. A sliding window of size (96,96,96) for MMWHS, (128,128,128) for MSD-BraTS and (32,32,32) for Hippocampus dataset, all with 25\% overlap was used for the inference. The experiments were conducted with PyTorch v1.12 and MONAI v0.9 framework \cite{monaicite} using a NVIDIA Quadro RTX 6000 GPU.

\subsection{Evaluation Metrics}
\label{subsec:evalmetrics}
Quantitative evaluations between predicted and ground truth segmentation regions were performed using the following evaluation metrics:
\begin{enumerate}
    \item \textbf{Dice score (Dice)}: The Dice similarity coefficient to measure the overlap between the GT Labels and the prediction from the network is given by:
    \begin{equation}
    \small
    Dice(G,Y) = \frac{2 \sum_{p=1}^N G_{p}Y_{p}}{\sum_{p=1}^N G_{p} + \sum_{p=1}^N Y_{p}}
    \label{equation:diceevalmetrics}
    \end{equation}
    where, $G_{p}$ and $Y_{p}$ denote the GT Labels and predicted output at voxel $p$. $N$ denotes the total number of voxels in the GT Labels and the predicted output. A larger Dice score indicates a better overlap between GT Labels and predicted segmentations.

    \item \textbf{Hausdorff distance (HD)}: Hausdorff Distance quantifies the dissimilarity between the boundary points in the ground truth labels and the predicted segmentation labels and is given by:
    \begin{equation}\label{hdevalmetrics}
    \small
    \begin{aligned}
        HD(G',Y') = \max \Big\{\max_{g' \in G'} \min_{y' \in Y'} \|g' - y'\|, \\
        \max_{y' \in Y'} \min_{g' \in G'} \|y' - g'\| \Big\} 
    \end{aligned}
    \end{equation}
    where, $G'$ and $Y'$ denote the GT surface point set and predicted surface point set, respectively. A smaller HD indicates a better agreement between GT and predicted surface point sets. The $95^{th}$ percentile of Hausdorff distance was used for all our experiments to reduce the effect of outliers.
\end{enumerate}

\begin{table*}[htbp]
\centering
\caption{Quantitative comparison for brain tumor segmentation task of MSD dataset with the DSD framework attached to UNETR, nnU-Net and Swin UNETR backbones. The mean Dice score (\%) and HD95 (mm) on the testing set are shown for each tumor structure. The highest Dice score and lowest HD95 marked in \textbf{bold}.}
\label{tab:Table2}
\resizebox{\textwidth}{!}{%
\begin{tabular}{c|cc|cc|cc|cc}
\hline
Anatomy & \multicolumn{2}{c|}{Whole Tumor (WT)} & \multicolumn{2}{c|}{Enhancing Tumor (ET)} & \multicolumn{2}{c|}{Tumor Core (TC)} & \multicolumn{2}{c}{Mean}            \\ \hline
Network   & Dice $\uparrow$  & HD95 $\downarrow$ & Dice $\uparrow$   & HD95 $\downarrow$   & Dice $\uparrow$ & HD95 $\downarrow$ & Dice $\uparrow$ & HD95 $\downarrow$ \\ \hline
UNet \cite{ronneberger2015u}                & 76.6 & 9.2  & 56.1 & 11.1 & 66.5 & 10.2 & 66.4          & 10.2         \\ 
AttUNet \cite{oktay2018attention}           & 76.7 & 9.0  & 54.3 & 10.4 & 68.3 & 10.5 & 66.5          & 10.0         \\ 
CoTr \cite{xie2021cotr}                     & 74.6 & 9.2  & 55.7 & 9.4  & 74.8 & 10.4 & 68.3          & 9.7          \\ 
TransBTS \cite{wang2021transbts}            & 77.9 & 10.0 & 57.4 & 10.0 & 73.5 & 9.0  & 69.6          & 9.7          \\ 
Light UNETR \cite{hatamizadeh2022unetr}     & 67.6 & 48.4 & 56.2 & 18.8 & 81.4 & 14.8 & 68.4          & 27.3         \\ 
Light UNETR with DSD (Ours)                 & 74.6 & 29.9 & 58.7 & 13.5 & 82.8 & 10.9 & 72.0          & 18.1         \\ 
UNETR \cite{hatamizadeh2022unetr}           & 75.2 & 22.6 & 53.6 & 9.8  & 78.1 & 14.8 & 68.9          & 15.7         \\ 
UNETR with DSD (Ours)                       & 80.4 & 9.8  & 64.1 & 8.0  & 85.2 & 3.6  & 76.6          & \textbf{7.1} \\ 
nnU-Net \cite{isensee2021nnu}               & 75.7 & 25.7 & 65.1 & 18.8 & 81.8 & 10.9 & 74.2          & 18.5         \\ 
nnU-Net with DSD (Ours)                     & 78.5 & 19.0 & 67.8 & 15.7 & 84.4 & 9.6  & \textbf{76.9} & 14.8         \\ 
Light Swin UNETR \cite{hatamizadeh2022swin} & 73.7 & 24.0 & 62.7 & 14.6 & 80.6 & 14.3 & 72.4          & 17.6         \\ 
Light Swin UNETR with DSD (Ours)            & 77.8 & 24.0 & 67.5 & 12.0 & 84.1 & 3.8  & 76.5          & 13.2         \\ \hline
\end{tabular}%
}
\end{table*}

\subsection{Prediction with MMWHS dataset}
\label{subsec:MMWHS}
Table \ref{tab:Table1} present the 5-fold cross validation results and summarizes the mean Dice score and HD95 of the 7 classes of cardiac substructures for the CT angiography (CTA) MMWHS dataset using UNETR and nnU-Net as the backbone along with the DSD framework. On average, across the 5-folds, when our proposed DSD framework is attached to the UNETR architecture, it outperforms the UNETR backbone with 2.7\% increase in the mean Dice score and 3.22 mm decrease in mean HD95. On average, across the 5-folds, when our proposed DSD framework is attached to the nnU-Net architecture, it outperforms the nnU-Net backbone with 2.94\% increase in the mean Dice score and 11.08 mm decrease in mean HD95. The improvements for both UNTER and nnU-Net with DSD attached to them, were achieved with negligible increase (less than 0.05\%) in the number of training parameters and training time per epoch. In Table \ref{tab:Table1}, we observe that the Dice score slightly drops for the MYO sub-structure when using nnU-Net with DSD. However, we also notice that there is a significant improvement in the HD95 for MYO when using nnU-Net with DSD. Since, HD95 is a measure of boundary distance between the GT and the predictions, it is clear that the basic nnU-Net (without DSD) has several isolated False predictions which leads to an increased boundary distance with the GT Labels. We also observe a significantly higher Dice score and lower HD95 for the complex and smaller sub-structures: Ascending Aorta (AA) and Pulmonary Artery (PA), which are clinically important and complex cardiac sub-structures that are harder to classify with a basic U-shaped network. 

The results with the MMWHS dataset show that our DSD framework could still capture the intricate patterns in the CTA images and improve the segmentation performance over the U-shaped backbones, even when we have an anisotropic dataset and a high slice thickness along the axial direction. This confirms that our approach could be applied to any U-shaped network and it works well with both isotropic and anisotropic datasets. 

Additionally, as shown in Table \ref{tab:Table1}, DSD also reduced the variance of the mean Dice score and mean HD95 across the 5-folds. A qualitative comparison (both on a 2D axial slice and 3D volume) between the GT Labels and prediction with UNETR backbone, UNETR with DSD, nnU-Net backbone and nnU-Net with DSD is shown in Fig. \ref{fig:figure2}. We deliberately showed different poses for UNETR and nnU-Net, as the improvements after incorporating DSD to the backbones, could be better seen from those poses. 

\begin{figure*}[ht]
\centering
\includegraphics[width=\textwidth]{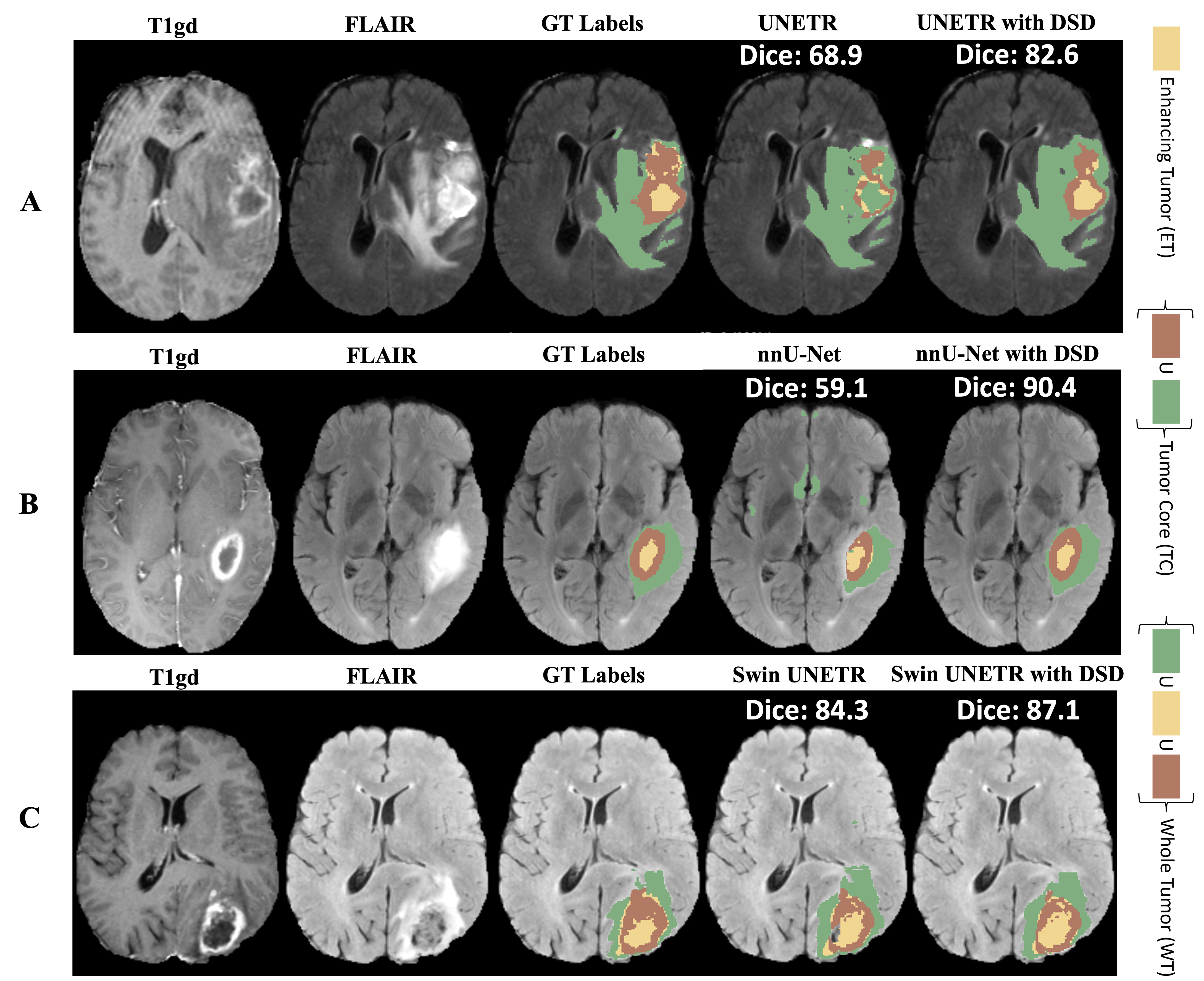}
\caption{Qualitative comparison of an axial slice with GT Labels (on FLAIR MRI) and predictions from (A) UNETR, (B) nnU-Net and (C) Swin UNETR, highlighting the improvements in segmentation (shown by Dice score (\%)) with our proposed DSD framework. The Dice score (\%) is for the full 3D volume of a patient belonging to the testing set.}
\label{fig:figure3}
\vspace{-1mm}
\end{figure*}

\begin{table*}[htbp]
\centering
\caption{Quantitative Comparison for Hippocampus segmentation task of MSD dataset with the DSD framework attached to VNet, UNETR and nnU-Net backbones. The mean Dice score (\%) and HD95 (mm) on the testing set are shown for each segmentation class. The highest Dice score and lowest HD95 are marked in \textbf{bold}. The number of training parameters (in millions) and training time (in minutes per epoch) is also shown.}
\label{tab:Table3}
\resizebox{\textwidth}{!}{%
\begin{tabular}{c|cc|cc|cc|c|c}
\hline
Anatomy &
  \multicolumn{2}{c|}{Anterior} &
  \multicolumn{2}{c|}{Posterior} &
  \multicolumn{2}{c|}{Mean} &
  \multirow{2}{*}{\# params} &
  \multirow{2}{*}{Time} \\ \cline{1-7}
Network &
  Dice $\uparrow$ &
  HD95 $\downarrow$ &
  Dice $\uparrow$ &
  HD95 $\downarrow$ &
  Dice $\uparrow$ &
  HD95 $\downarrow$ &
   &
   \\ \hline
VNet \cite{milletari2016v} &
  83.95 $\pm$ 7.92 &
  2.19 $\pm$ 1.53 &
  83.68 $\pm$ 4.95 &
  3.59 $\pm$ 4.51 &
  83.82 $\pm$ 6.58 &
  2.89 $\pm$ 3.43 &
  45.602 &
  0.26 \\ 
\begin{tabular}[c]{@{}c@{}}VNet with DSD (Ours)\end{tabular} &
  85.54 $\pm$ 6.61 &
  1.55 $\pm$ 1.08 &
  85.36 $\pm$ 3.50 &
  2.08 $\pm$ 2.71 &
  85.45 $\pm$ 5.05 &
  1.82 $\pm$ 1.89 &
  45.605 &
  0.29 \\ 
UNETR \cite{hatamizadeh2022unetr} &
  84.36 $\pm$ 4.38 &
  3.53 $\pm$ 4.62 &
  84.29 $\pm$ 3.74 &
  1.92 $\pm$ 0.93 &
  84.32 $\pm$ 4.05 &
  2.73 $\pm$ 3.42 &
  92.624 &
  0.17 \\ 
\begin{tabular}[c]{@{}c@{}}UNETR with DSD (Ours)\end{tabular} &
  85.85 $\pm$ 3.21 &
  1.95 $\pm$ 0.85 &
  85.38 $\pm$ 3.48 &
  1.85 $\pm$ 0.87 &
  85.62 $\pm$ 3.34 &
  1.90 $\pm$ 0.86 &
  92.628 &
  0.21 \\ 
nnU-Net \cite{isensee2021nnu} &
  86.49 $\pm$ 5.27 &
  2.38 $\pm$ 4.04 &
  85.84 $\pm$ 3.52 &
  1.65 $\pm$ 0.85 &
  86.16 $\pm$ 4.47 &
  2.02 $\pm$ 2.93 &
  5.6870 &
  0.15 \\ 
\begin{tabular}[c]{@{}c@{}}nnU-Net with DSD (Ours)\end{tabular} &
  87.34 $\pm$ 4.34 &
  1.64 $\pm$ 0.66 &
  86.86 $\pm$ 3.40 &
  1.62 $\pm$ 0.70 &
  \textbf{87.10} $\pm$ 3.90 &
  \textbf{1.63} $\pm$ 0.68 &
  5.6890 &
  0.17 \\ \hline
\end{tabular}
}
\end{table*}

\subsection{Prediction with Brain Tumor dataset}
\label{subsec:brats_data}
Table \ref{tab:Table2} summarizes the prediction results for the brain tumor segmentation task of the MSD dataset. Our DSD framework is attached to four different U-shaped backbone architectures and compared to several state-of-the-art architectures on the MSD-BraTS dataset. We attached our DSD framework on a light UNETR (with feature size 16) as well as a standard UNETR (with feature size 64) backbone. When the DSD framework was attached to the light UNETR backbone, it outperforms the backbone with 3.6\% increase in mean Dice score and 9.2 mm decrease in mean HD95. Additionally, when the DSD framework was attached to a standard UNETR (with feature size 64) backbone, it outperforms the backbone with 7.7\% increase in mean Dice score and 8.6 mm decrease in mean HD95. Interestingly, the light UNETR with DSD even outperforms the standard UNETR with a 3.1\% increase in mean Dice score but with 39.7\% less training parameters.  

Table \ref{tab:Table2} also shows that when the DSD framework is attached to the nnU-Net backbone, it outperforms the backbone with 2.7\% increase in the mean Dice score and 3.7 mm decrease in mean HD95.

Additionally, as observed in Table \ref{tab:Table2}, when the DSD framework is attached to a light Swin UNETR backbone (with feature size 12), it outperforms the backbone with 4.1\% increase in the mean Dice score and 4.4 mm decrease in mean HD95. The improvements for UNTER, nnU-Net and Swin UNETR with DSD attached to them, were achieved with negligible increase (less than 0.07\%) in the number of training parameters and training time per epoch.

\begin{figure}[htbp]
\centering
\includegraphics[width=0.7\textwidth]{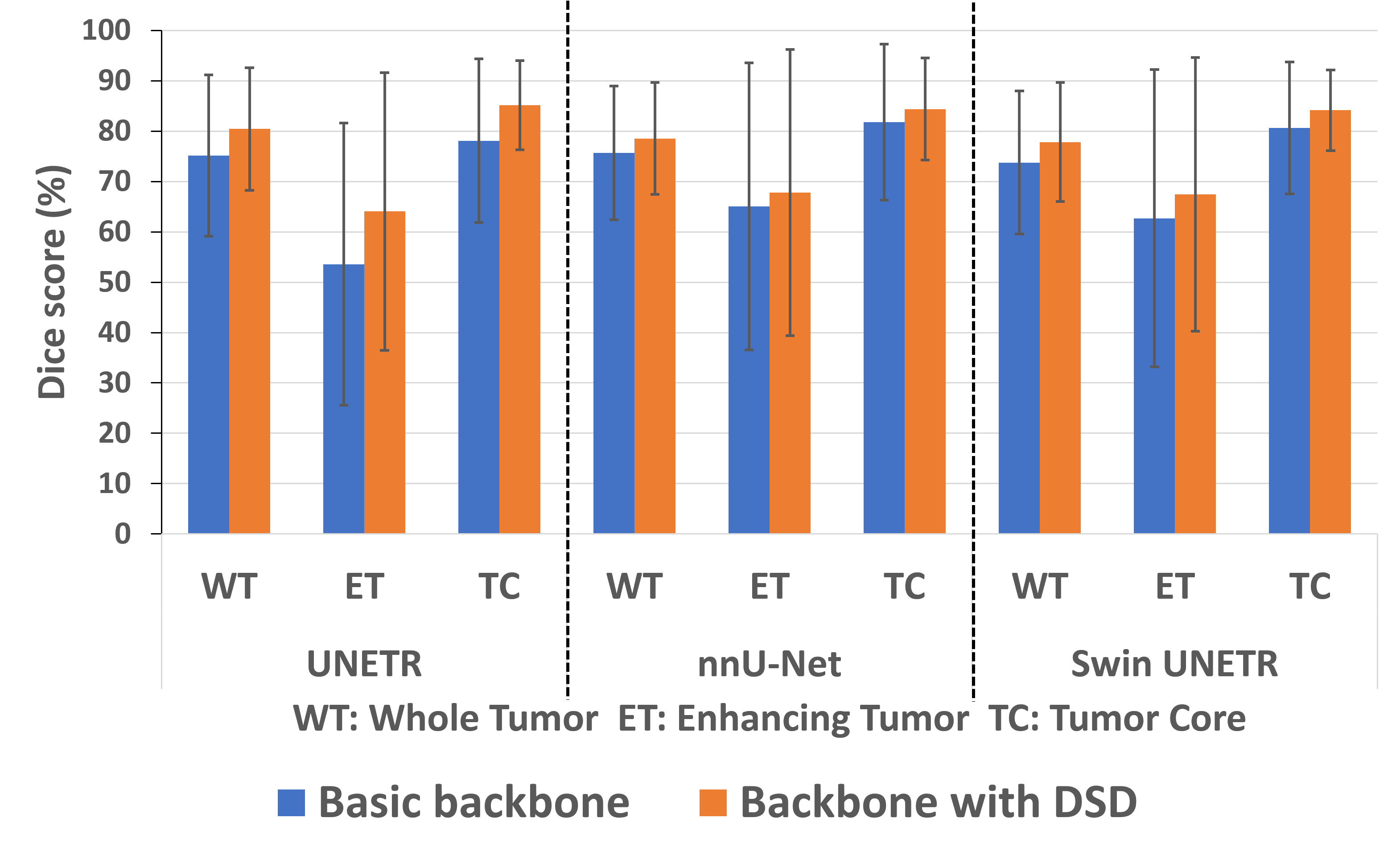}
\caption{The mean and standard deviation of Dice score (\%) for all patients in the testing set for each foreground class of MSD-BraTS.}
\label{fig:figure4}
\vspace{-4mm}
\end{figure}

We observe from Table \ref{tab:Table2} that the standard UNETR with DSD and nnU-Net with DSD achieves the best performance when compared with the state-of-the-art methods on the MSD-BraTS dataset \cite{ronneberger2015u,oktay2018attention,xie2021cotr,wang2021transbts,hatamizadeh2022unetr,isensee2021nnu,hatamizadeh2022swin}. A qualitative comparison between the GT Labels and predictions with UNETR backbone, UNETR with DSD, nnU-Net backbone, nnU-Net with DSD, Swin UNETR backbone and Swin UNETR with DSD on a 2D slice (on FLAIR MRI) is shown in Fig. \ref{fig:figure3}. When using the DSD framework with the U-shaped backbones on the MSD-BraTS dataset, we observed an increased mean and reduced variance of the Dice score of each class for all patients in the testing set as shown in Fig. \ref{fig:figure4}. 

\subsection{Prediction with Hippocampus dataset}
\label{subsec:hippocampus}
Table \ref{tab:Table3} summarizes the prediction results for the Hippocampus segmentation task of the MSD dataset. Our DSD framework is attached to three different U-shaped backbone architectures on the MSD-Hippocampus dataset. When the DSD framework was attached to the VNet backbone, it outperforms the backbone with 1.63\% increase in mean Dice score and 1.07 mm decrease in mean HD95. 

Table \ref{tab:Table3} also shows that when the DSD framework is attached to the UNETR backbone, it outperforms the backbone with 1.3\% increase in the mean Dice score and 0.83 mm decrease in mean HD95. Additionally, when attached to the nnU-Net backbone, DSD outperforms the backbone with 0.94\% increase in the mean Dice score and 0.39 mm decrease in mean HD95. Given the low resolution of the Hippocampus volumes, these are significant improvements in the performance.  

\begin{figure}[htbp]
\centering
\includegraphics[width=0.99\columnwidth]{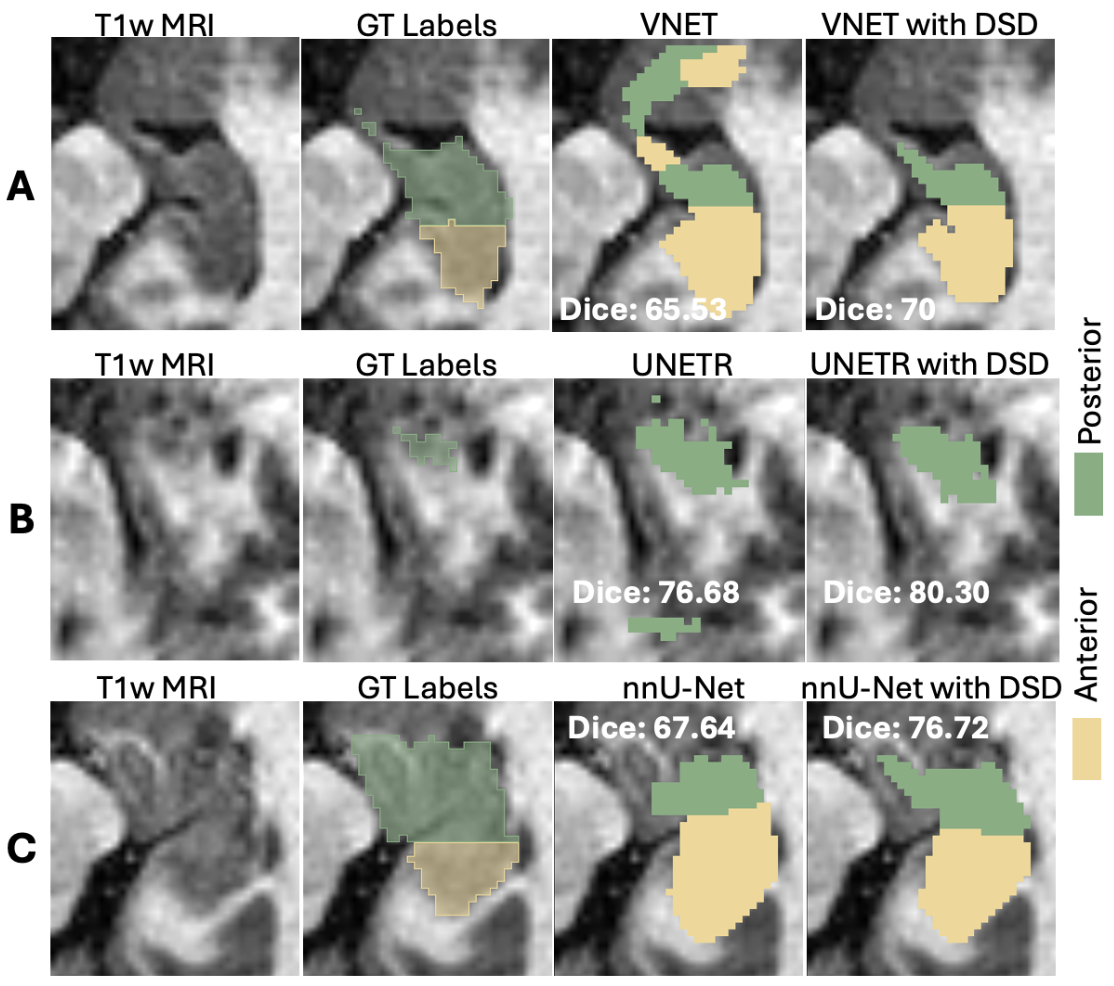}
\caption{Qualitative comparison of an axial slice with GT Labels (on T1w MRI) and predictions from (A) VNET, (B) UNETR and (C) nnU-Net, highlighting the improvements in segmentation (shown by Dice score (\%)) with our proposed DSD framework. The Dice score (\%) is for the full 3D volume of a patient belonging to the testing set.}
\label{fig:figure5}
\vspace{-1mm}
\end{figure}

The improvements for VNet, UNTER and nnU-Net with DSD attached to them, were achieved with negligible increase in the number of training parameters (less than 0.03\%) and training time per epoch. When using the DSD framework with the U-shaped backbones on the MSD-Hippocampus dataset, we observed an increased mean and reduced variance of the Dice score and HD95 of each class for all patients in the testing set as shown in Table \ref{tab:Table3}. A qualitative comparison between the GT Labels and predictions with VNet backbone, VNet with DSD, UNETR backbone, UNETR with DSD, nnU-Net backbone and nnU-Net with DSD on a 2D axial slice (on T1-w MRI) is shown in Fig. \ref{fig:figure5}. We would like to clarify that the Dice score reported in Fig. \ref{fig:figure5} pertains to the entire 3D volume, whereas only one 2D slice is depicted. In Row C, the improvement is distributed in smaller increments across many slices, leading to a sizable 9.08\% overall gain. Conversely, for Rows A and B, we deliberately selected slices that visually accentuate the segmentation differences. Thus, even if Row C’s selected slice does not reveal a dramatic change, the collective effect across all slices in the volume contributes to the higher Dice score.

\begin{table*}[htbp]
\centering
\caption{Comparison of Basic UNETR, UNETR with MISSU and UNETR with DSD on segmentation performance shown by the mean Dice score (\%) of all 7 cardiac substructures with one fold of MMWHS validation set and the Hippocampus testing set. The number of training parameters (in Millions) and training time (in minutes per epoch) for all three networks have been shown.}
\label{tab:Table4}
\resizebox{0.9\textwidth}{!}{%
\begin{tabular}{c|ccc|ccc}
\hline
\multirow{2}{*}{Network}                                 & \multicolumn{3}{c|}{MMWHS} & \multicolumn{3}{c}{Hippocampus} \\ \cline{2-7} 
                      & Dice $\uparrow$ & \# params & Time & Dice $\uparrow$ & \# params & Time \\ \hline
Basic UNETR \cite{hatamizadeh2022unetr} & 76.12  & 101.787  & 2.15  & 84.32     & 92.624    & 0.17    \\ \hline
UNETR with MISSU \cite{wang2023missu}   & 78.05  & 114.353  & 2.22  & 84.00     & 97.094    & 0.25    \\ \hline
UNETR with DSD (ours) & 80.93           & 101.804   & 2.16 & 85.62           & 92.628    & 0.21 \\ \hline
\end{tabular}
}
\end{table*}

\subsection{Comparison with MISSU on U-shaped backbone}
\label{subsec:missu}
MISSU \cite{wang2023missu} is another self-distillation framework that was proposed to improve 3D medical image segmentation performance. Table \ref{tab:Table4} compares the quantitative prediction results of the Basic UNETR \cite{hatamizadeh2022unetr} with MISSU and DSD attached to it. For the UNETR with MISSU, we implemented the multi-scale fusion (MSF) blocks as stated in \cite{wang2023missu} and then attached it to the UNETR architecture to perform self-distillation (with KL-divergence loss) between the intermediate encoder outputs and the MSF block outputs. We observe that the UNETR with our proposed DSD framework outperforms the UNETR with MISSU by a significant margin. Specifically, we observe a 2.88\% and 1.62\% improvement in Dice scores when DSD is attached to UNETR compared to MISSU attached to UNETR, for MMWHS and Hippocampus datasets respectively. We also observe that these improvements were achieved by DSD with significantly less number of training parameters compared to MISSU.

\begin{table}[htbp]
\centering
\caption{Ablation study showing the mean Dice score of all 7 cardiac substructures of one fold of MMWHS validation set obtained by different components of our DSD framework. The highest Dice score (\%) is marked in \textbf{bold}.}
\label{tab:Table5}
\resizebox{0.6\textwidth}{!}{%
\begin{tabular}{c|c|c|c|c|c}
\hline
Study settings & Basic & DS    & SDD   & SDE   & DSD                             \\ \hline
Network        & Dice$\uparrow$  & Dice$\uparrow$  & Dice$\uparrow$  & Dice$\uparrow$  & Dice$\uparrow$                           \\ \hline
UNETR          & 76.12 & 78.29 & 80.13 & 79.90 & \textbf{80.93} \\ \hline
nnU-Net        & 76.97 & 83.74 & 84.02 & 84.42 & \textbf{86.10} \\ \hline
\end{tabular}%
}
\end{table}

\subsection{Ablation Study}
\label{subsec:ablation}
Table \ref{tab:Table5} presents the results of an ablation study with both UNETR and nnU-Net on one fold of the MMWHS validation set comparing the following components of our DSD framework: (i) \textbf{Basic} ($\eta = 0$ and $\alpha_1,\alpha_2 = 0$ in Eq. \ref{equation:KL_loss_final}): This is the basic U-shaped backbone without any deep supervision and self-distillation. (ii) \textbf{DS} (Deep Supervision only: $\eta = 1$ and $\alpha_1,\alpha_2 = 0$ in Eq. \ref{equation:KL_loss_final}): DSD in this case is reduced to deep supervision where the knowledge is distilled from the GT Labels to the deeper and shallower decoder layers of the U-shaped backbone. DS improves the Dice score by 2.17\% over basic UNETR and 6.77\% over basic nnU-Net. (iii) \textbf{SDD} (Self-Distillation between Decoders only: $\eta = 1$, $\alpha_1 = 0$, $\alpha_2 = 1$ in Eq. \ref{equation:KL_loss_final}): This shows the effect of self-distillation only between the decoder layers along with deep supervision in DSD. SDD improves the Dice score by 1.84\% over UNETR with DS and 0.28\% over nnU-Net with DS. (iv) \textbf{SDE} (Self-Distillation between Encoders only: $\eta = 1$, $\alpha_1 = 1$, $\alpha_2 = 0$ in Eq. \ref{equation:KL_loss_final}): This shows the effect of self-distillation only between the encoder layers along with deep supervision in DSD. SDE improves the Dice score by 1.61\% over UNETR with DS and 0.68\% over nnU-Net with DS. (v) \textbf{DSD} ($\eta = 1$, $\alpha_1 = 1$, $\alpha_2 = 1$ in Eq. \ref{equation:KL_loss_final}): This shows the effect of our proposed DSD where knowledge distills from the deeper to the shallower layers both on the encoder and decoder side of the U-shaped backbone. DSD achieves the best performance with 2.64\% and 2.36\% increase in Dice score over UNETR with DS and nnU-Net with DS, respectively. We also performed quantitative studies on the choice of the temperature hyperparameter ($\tau$ in Eq. \ref{equation:softlabelsgenerator}) for all our experiments. When we attached DSD to the nnU-Net architecture with $\tau$ = 1, 2, 3 and 4, we observed the Dice scores to be 85.37\%, 85.89\%, 86.10\% and 86.07\% respectively, on one-fold of the MMWHS validation set.

The reason behind the improvement in performance with the different study settings in Table \ref{tab:Table5} is as follows: DS improves upon the Basic model by directly supervising intermediate decoder layers, forcing them to predict the segmentation mask and thereby mitigating vanishing gradients while promoting early learning of discriminative features. In SDD, the deepest decoder—with its refined semantic predictions—serves as a teacher to guide shallower decoders, ensuring consistency in feature representations and predictions. Similarly, in SDE, the deepest encoder, which captures rich contextual information, guides the shallower encoders to extract more robust features. DSD combines these strategies, simultaneously enhancing feature extraction in the encoder and prediction consistency in the decoder, leading to improved regularization, better gradient flow, and a more coherent network representation.

\begin{table}[htbp]
\centering
\caption{Ablation study comparing the mean Dice score of Hippocampus and MMWHS dataset on the validation set, obtained using the Basic, Mean Teacher, Reversed Teacher and our DSD framework. The highest Dice score (\%) is marked in \textbf{Bold}.}
\label{tab:Table6}
\resizebox{0.6\textwidth}{!}{%
\begin{tabular}{c|c|c|c|c|c}
\hline
Dataset & Model & Basic & \begin{tabular}[c]{@{}c@{}}Mean \\ Teacher\end{tabular} & \begin{tabular}[c]{@{}c@{}}Reversed \\ Teacher\end{tabular} & DSD \\ \hline
Hippocampus & UNETR   & 85.50 & 87.06 & 86.39 & \textbf{87.71} \\ \hline
Hippocampus & nnU-Net & 88.84 & 89.23 & 88.97 & \textbf{89.86} \\ \hline
MMWHS       & UNETR   & 76.12 & 78.92 & 77.23 & \textbf{80.93} \\ \hline
MMWHS       & nnU-Net & 76.97 & 77.79 & 77.15 & \textbf{86.10} \\ \hline
\end{tabular}
}
\end{table}

\begin{table*}[!ht]
\centering
\caption{Effect of network size (number of parameters) on segmentation performance shown by the mean Dice score (\%) of all 7 cardiac substructures with one fold of MMWHS validation set obtained by varying the network size. The number of training parameters (in Millions) and training time (in minutes per epoch) for both the Small and Large Networks have been shown.}
\label{tab:Table7}
\resizebox{\textwidth}{!}{%
\begin{tabular}{c|ccc|ccc|ccc|ccc}
\hline
\begin{tabular}[c]{@{}c@{}}Network \\ size\end{tabular} &
  \multicolumn{3}{c|}{\begin{tabular}[c]{@{}c@{}}Small Network \\ (No DSD)\end{tabular}} &
  \multicolumn{3}{c|}{\begin{tabular}[c]{@{}c@{}}Small Network \\ (DSD)\end{tabular}} &
  \multicolumn{3}{c|}{\begin{tabular}[c]{@{}c@{}}Large Network \\ (No DSD)\end{tabular}} &
  \multicolumn{3}{c}{\begin{tabular}[c]{@{}c@{}}Large Network \\ (DSD)\end{tabular}} \\ \hline
Network & Dice $\uparrow$ & \# params & Time & Dice $\uparrow$ & \# params & Time & Dice $\uparrow$ & \# params & Time & Dice $\uparrow$ & \# params & Time \\ \hline
UNETR   & 72.91           & 92.784    & 2.12 & 78.88           & 92.797    & 2.13 & 76.12           & 101.787   & 2.15 & 80.93           & 101.804   & 2.16 \\ \hline
nnU-Net & 74.89           & 5.6870    & 2.05 & 79.43           & 5.6950    & 2.06 & 76.97           & 16.6640   & 2.05 & 86.10           & 16.6720   & 2.09 \\ \hline
\end{tabular}
}
\vspace{-1mm}
\end{table*}

We have also performed additional ablation studies on one fold of the MMWHS validation set and the Hippocampus validation set as shown in Table \ref{tab:Table6}. The rationale behind these experiments is to validate our design choice for DSD, i.e., to demonstrate that we are correctly distilling knowledge for both encoders and decoders in our DSD framework. The Basic ((i) above) and DSD ((v) above) have the same settings as shown in Table \ref{tab:Table5}. In the Mean Teacher experiment, the mean of all encoder outputs was used as teacher for all individual encoder outputs (students), and the mean of all decoder outputs was used as teacher for all individual decoder outputs (students) in the DSD process of the U-shaped networks. We observe that with this setting, there is a significant drop in the dice score. For the Hippocampus dataset, we observed a 0.65\% drop for UNETR and a 0.63\% drop for the nnU-Net, compared to DSD. For the MMWHS dataset, we observed a 2.01\% drop for UNETR and an 8.31\% drop for the nnU-Net, compared to DSD. In the Reversed Teacher experiment, we have reversed the direction of knowledge transfer on the encoder and decoder sides in our DSD framework, i.e., the knowledge was transferred in a top-down manner on the encoder side (from the shallowest encoder (Encoder 1) to the deeper encoders (Encoder 2, …, Encoder Z)) and in a bottom-up manner on the decoder side (from the shallowest decoder (Decoder Z) to the deeper decoders (Decoder 1, …, Decoder (Z-1))) [See Fig. \ref{fig:figure1}]. For the Hippocampus dataset, we observed a 1.32\% drop for UNETR and a 0.89\% drop for the nnU-Net. For the MMWHS dataset, we observed a 3.7\% drop for UNETR and an 8.95\% drop for the nnU-Net.

The reason behind the poor performance of the Mean Teacher and Reversed Teacher method in Table \ref{tab:Table6} is as follows: In the Mean Teacher experiment, the representations for the teacher are smoothed-out due to the averaging operation (mean of all encoder or decoder representations), which produces a smoothed teacher signal that dilutes the refined, high-level details present in the deepest layers. This smoothing effect weakens the guidance provided to individual student layers, reducing the network’s ability to learn rich, context-specific features, and hence the Mean Teacher model underperformed our DSD approach. In the Reversed Teacher experiment, the performance drop is due to the fact that we are reversing the direction of knowledge transfer between the encoders and decoders of the U-shaped network. This reverses the natural information flow: in the encoder, shallower layers (with limited context) incorrectly guide deeper layers, and in the decoder, early-stage (less semantic) predictions teach later layers. This inversion undermines the hierarchical feature extraction and semantic refinement that are essential for accurate segmentation. The above experiments show that the design choices made on both the encoder and decoder side in our DSD approach are appropriate.

The original self-distillation implementation used the L2 loss between the feature maps of the teacher and student \cite{zhang2019your,zhang2021self}. In our ablation study, we also applied the L2 loss to the self-distillation between the feature maps of encoders and decoders. However, no noticeable improvement in performance was observed with this strategy, likely because we are performing pixel-level classification where the loss (Eq. \ref{equation:KL_loss_final}) from each pixel is back-propagated to update the network weights, and hence the loss from feature maps become redundant. This is different from the image-level classification tasks \cite{zhang2019your,zhang2021self}, where the loss from feature maps help improve the classification performance. Thus, feature map-based distillation was not included in the rest of our experiments.

\subsection{Effect of parameters on network performance}
\label{subsec:params}
Table \ref{tab:Table7} presents the effect of the number of trainable parameters on the performance of UNETR and nnU-Net on one fold of the MMWHS validation set. A small UNETR architecture with a feature size of 16 is compared with a large UNETR architecture with a feature size of 32. All the other settings were the same as the large UNETR, described in Section \ref{subsec:expsetup}. When a small UNETR with less trainable parameters is used, applying the DSD to this small network improves the Dice score by 5.97\% with only 0.014\% increase in the number of trainable parameters and with negligible increase in training time per epoch. On the other hand, a large UNETR only leads to 3.21\% increase in Dice score when compared with a small UNETR although there was a 9.7\% increase in the number of trainable parameters with more training time per epoch. Finally, using DSD on the large UNETR improves the Dice score further by 4.81\% over the large UNETR without DSD. Similarly, we compared a small nnU-Net with 16 and 256 filters for the input and bottleneck blocks and 32, 64 and 128 for both the downsample and upsample blocks. The other settings were the same as a large nnU-Net as described in Section \ref{subsec:expsetup}. It should be noted that for the nnU-Net architecture, we only changed the initial setting for the number of filters for large and small networks such that the “Large Network” is initialized with much more filters while keeping all other hyperparameters the same. That is, the kernel sizes, strides and number of layers are all initialized the same for both “Small Network” and “Large Network”. Applying the DSD to a small nnU-Net improves the Dice score by 4.54\% with only 0.14\% increase in the number of parameters. In contrast, a large nnU-Net increases the Dice score by only 2.08\% over the smaller nnU-Net although the number of trainable parameters increases by 193\% with significant increase in training time per epoch. Finally, a large nnU-Net with DSD improves the Dice score further by 9.13\% over the large nnU-Net without DSD, demonstrating that a nnU-Net initialized with a larger number of filters cannot match the performance of a nnU-Net with DSD initialized with a smaller number of filters. 

In summary, the observations presented in Table \ref{tab:Table7} highlight that attaching the DSD framework to the U-shaped architecture can significantly improve the prediction results with only slight increases on the number of trainable parameters, brought by the bottleneck modules. We used the large networks with DSD in the rest of our experiments since they yielded the best performance. 

\section{Discussion}
\label{sec:discussion}
The present study investigated the application of self-distillation for medical image segmentation. The goal of this study was to understand whether self-distillation could further improve the segmentation performance of a single U-shaped network to its maximum extent possible by utilizing the knowledge of the deepest encoder and decoder in addition to the GT Labels, by using very few additional trainable parameters and without the need of a large pretrained model as in the case of knowledge distillation. Minimizing the KL-divergence in the loss function between the deepest encoder/decoder and shallower encoders/decoders has a regularization effect which helps the model generalize better to unseen data.

All of our experiments were performed with the same initialization strategy to ensure that the boost in the segmentation performance when using the DSD framework is by leveraging the regularization effect of the KL divergence loss between encoders and decoders and not due to the random initialization of the network layers.  
We demonstrated that just increasing the number of trainable parameters did not boost the segmentation performance by an amount similar to when the DSD framework is attached to the backbones. Thus, DSD was shown to significantly boost the segmentation performance of a light-weight backbone and outperform  full size models. This shows that DSD is indeed an efficient training strategy to enhance the medical image segmentation performance. 

The observed discrepancies in the performance improvements of our proposed DSD framework across different datasets (MMWHS, MSD-BraTS, and MSD-Hippocampus) and backbone networks are expected due to inherent differences in dataset complexity, modality, and anatomical structures. For instance, the MSD-BraTS multi-modal MRI dataset exhibits high variability in tumor morphology and complexity, making the segmentation inherently more challenging and hence leading to greater potential improvements with DSD attached, compared to simpler, single-modality datasets such as MSD-Hippocampus with T1w MRI and MMWHS CT datasets. These variations illustrate the broad applicability and effectiveness of our DSD framework. Additionally, in Table \ref{tab:Table4}, the observed discrepancies in performance stem from the varying anatomical complexity, size, and image characteristics of each dataset. In MMWHS (CT modality), multiple large cardiac substructures with clearer boundaries create more opportunities for inter- and intra-structure knowledge transfer, leading to pronounced gains with the DSD framework. By contrast, the Hippocampus dataset (T1w MRI) presents smaller, more localized structures with subtler intensity gradients, limiting the available learning signal and reducing the magnitude of improvement. Consequently, while the DSD framework consistently benefits all tested datasets, the degree of enhancement depends on these factors, explaining the stronger results in MMWHS and more modest gains for Hippocampus.

The most closely related prior work to DSD is MISSU \cite{wang2023missu}, which applies self-distillation to a transformer-based TransUNet \cite{chen2021transunet}, incorporating multi-scale fusion blocks for 3D medical image segmentation. It is worth mentioning that our DSD framework has several advantages over MISSU: (i) DSD is highly adaptable and designed to be compatible with any U-shaped backbone (including VNet, nnU-Net, UNETR, Swin UNETR), while MISSU is tightly integrated with the TransUNet, thereby making it less versatile. (ii) DSD provides a more thorough internal knowledge transfer involving two complementary pathways (encoder and decoder-side distillation), while MISSU relies on a single-path self-distillation process focused on multi-scale feature fusion. (iii) The added learnable bottleneck modules in DSD incur negligible increase in the number of trainable parameters and training time, while the series of atrous convolutions in MISSU significantly increases the number of trainable parameters. (iv) DSD alleviates gradient vanishing issues with deep supervision which helps ensure that shallow layers can benefit directly from both the ground truth and deeper layers, while MISSU does not have a direct supervision at multiple levels that can lead to slower convergence and potentially less effective gradient flow. (v) Quantitative experiments with the MISSU attached to the UNETR framework demonstrate that our DSD framework with UNETR is superior as it obtains a higher Dice score with a lower number of training parameters on both MMWHS and Hippocampus datasets. In summary, while both DSD and MISSU contribute valuable techniques for improving segmentation performance, DSD’s advantages lie in its flexibility, lower computational burden, and innovative DSD process. 

One of the limitation of our approach is its dependency on the U-shaped network structure which means the benefits may not directly translate to other architectures without significant adaptation. DSD also relies on a carefully chosen teacher–student pairing to deliver precise, high-quality guidance and in scenarios where the deepest layers do not provide optimal representations, the improvement might be limited. Additionally, we observed that there are some outlier predictions for some of the complex substructures even after DSD has been attached to the U-shaped backbone. Hence, in our future work, we plan to investigate the effect of shape priors from the GT Labels on the DSD framework to further improve our segmentation results, especially for smaller complex structures.

\section{Conclusion}
\label{sec:conclusion}
In this paper, we introduced a novel DSD framework which could be attached into any U-shaped backbone for volumetric medical image segmentation. We incorporated our DSD framework into UNETR, nnU-Net, VNet and Swin UNETR and evaluated it on various benchmark datasets with promising results. Our results demonstrated that DSD is an efficient and general training strategy that could further boost the segmentation performance of these U-shaped networks.

\end{document}